\documentstyle[12pt]{article}
\newcommand{\nz}{\ifmmode \menge{I\hskip -.2ex N} \else
                          {\sf I\hskip -.2ex N} \fi}
\newcommand{\zz}{\ifmmode \menge{Z\hskip -.9ex Z} \else
                          {\sf Z\hskip -.9ex Z} \fi}
\newcommand{\kz}{\ifmmode \menge{I\hskip -.2ex K} \else
                          {\sf I\hskip -.2ex K} \fi}
\newcommand{\qz}{\ifmmode \menge{Q\hskip -1.3ex I\hskip .7ex} \else
                          {\sf Q\hskip -1.3ex I\hskip .7ex} \fi}
\newcommand{\rz}{\ifmmode \menge{I\hskip -.2ex R} \else
                          {\sf I\hskip -.2ex R} \fi}
\newcommand{\cz}{\ifmmode \menge{C\hskip -1ex I\hskip .4ex} \else
                          {\sf C\hskip -1ex I\hskip .4ex} \fi}

\begin{document}

\title{Generation of arbitrary two dimensional motional states of a trapped ion}
\author{XuBo Zou, K. Pahlke and W. Mathis  \\
\\Institute TET, University of Hannover,\\
Appelstr. 9A, 30167 Hannover, Germany }
\date{}

\maketitle

\begin{abstract}
{\normalsize We present a scheme to generate an arbitrary
two-dimensional quantum state of motion of a trapped ion. This
proposal is based on a sequence of laser pulses, which are tuned
appropriately to control transitions on the sidebands of two modes
of vibration. Not more than $(M+1)(N+1)$ laser pulses are needed
to generate a pure state with
a phonon number limit $M$ and $N$.\\

}

\end{abstract}
The generation of nonclassical states was studied in the past
theoretically and experimentally. The first significant advances
were made in quantum optics by demonstrating antibunched light
\cite{Paul} and squeezed light \cite{Loudon}. Various optical
schemes for generating Schr\"{o}dinger cat states were studied
\cite{YSB}, which led to an experimental realization in a
quantized cavity field \cite{Brune}. Several schemes were proposed
to generate any single-mode quantum state of a cavity field
\cite{VAP,PMZK,LE} and traveling laser field \cite{DCKW}.
Recently, possible ways of generating various two-mode entangled
field states were proposed. For example, it was shown
\cite{Sanders} that entangled coherent states, which can be a
superposition of two-mode coherent states \cite{Sanders,Chai} can
be produced using the nonlinear Mach-Zehnder interferometer. These
quantum states can be considered as a two(multi)-mode
generalization of single-mode Schr\"{o}dinger cat states
\cite{Manko}. A method to generate another type of two-mode
Schr\"{o}dinger cat states, which are known as SU(2)
Schr\"{o}dinger cat states, was proposed \cite{Sanders2,GGJMO}.
These quantum states result when two different SU(2) coherent
states \cite{Buzek,Sanders2,GGJMO} are superposed. It was also
shown that two-mode entangled number states can be generated by
using nonlinear optical interactions \cite{Gerry}, which may then
be used to obtain the maximum sensitivity in phase measurements
set by the Heisenberg limit \cite{Bollinger} . In general,
however, an experimental realization of nonclassical field states
is difficult, because the quantum coherence can be destroyed
easily
by the interaction with the environment.\\
Recent advances in ion cooling and trapping have opened new
prospects in nonclassical state generation. An ion confined in an
electromagnetic trap can be described approximately as a particle
in a harmonic potential. Its center of mass (c.m.) exhibits a
simple quantum-mechanical harmonic motion. By driving the ion
appropriately with laser fields, its internal and external degrees
of freedom can be coupled to the extent that its center-of-mass
motion can be manipulated precisely. One advantage of the trapped
ion system is that decoherence effects are relatively weak due to
the extremely weak coupling between the vibrational modes and the
environment. It was realized that this advantage of the trapped
ion system makes it a promising candidate for constructing quantum
logic gates for quantum computation \cite{CZ} as well as for
producing nonclassical states of the center-of-mass motion. In
fact, single-mode nonclassical states, such as Fock states,
squeezed states and Schr\"{o}dinger cat states of the ion's
vibration mode were investigated \cite{CCVPZ,Gerry2,MM}. Recently,
various schemes of producing two-mode nonclassical states of the
vibration mode were proposed using ions in a two-dimensional trap
\cite{Gerry2,GN,GGM}. In particular, several schemes were proposed
to generate arbitrary two-dimensional motional quantum states of a
trapped ion
\begin{eqnarray}
\sum_{m=0}^{M}\sum_{n=0}^{N}C_{m,n}|m>_x|n>_y \, .\label{1}
\end{eqnarray}
The ket-vectors $|m>_x$ and $|n>_y$ denote the Fock states of the
quantized vibration along the $x$ and $y$ axes. In the schemes of
Gardiner et al \cite{GCZ}, the number of required laser operations
depends exponentially on the upper photon numbers $M$ and $N$. The
scheme introduced by Kneer and Law \cite{KL} involves
(2M+1)(N+1)+2N operations, while the scheme proposed by Drobny et
al \cite{DHB} requires $2(M+N)^2$ operations. More recently S. B.
Zheng presented a scheme \cite{ZH}, which requires only (M+2)(N+1)
operations. The reduction of the number of operation is important
for further experimental realization. Notice that the quantum
state (\ref{1}) involves (M+1)(N+1) complex coefficients. The
purpose of this paper is to present a scheme to generate quantum
state (\ref{1}) with not more than (M+1)(N+1) quantum operations.
We will show that each coefficient of quantum state (\ref{1})
corresponds to only one laser pulse of this most efficient quantum
state generation scheme. In order to describe this concept in an
obvious way, we split the Hilbert space $H_{vib}$ of the two modes
of vibration to subspaces $H_{vib}^{2J}$ with a constant total
number of vibrational quanta $m+n=2J$. Thus, the Hilbert-space is
a direct sum: $H_{vib}=\bigcup_{2J=0}^{\infty}H_{vib}^{2J}$ with
$2J\in$\nz. The quantum states are formulated by the two-mode Fock
states $|J,L>=|J+L>_x|J-L>_y$ with $L\in \{-J,-J+1,\cdots ,J\}$.
So $J$ and $L$ can be half numbers. The subspaces $H_{vib}^{2J}$
are spanned by $|J,L>;|J,L-1>;\ldots;|J,-L>$. The quantum state
(\ref{1}) can also be written in the form
\begin{eqnarray}
\sum_{J=0}^{(M+N)/2}\sum_{L=-J}^{J}d_{J,L}|J,L>\,. \label{2}
\end{eqnarray}
In order to synthesize 2D quantum states of the vibration of one
trapped ion we suggest to use laser stimulated Raman processes
\cite{Steinbach}. We consider a trapped ion confined in a 2D
harmonic potential characterized by the trap frequencies $\nu_x$
and $\nu_y$ in two orthogonal directions $x$ and $y$. The ion is
irradiated along the $x$ and $y$ axes by two external laser
frequencies $\omega_x$, $\omega_y$ and wave vectors $k_x$, $k_y$.
The two laser fields stimulate Raman transitions between two
electronic ground state levels $|g_1>$ and $|g_2>$ via a third
electronic level, which is far enough out of resonance. The
difference of energy of these two electronic ground states is
$\hbar \omega_0$. We concentrate our investigations on those
transitions on the quantum states of the trapped ion, which are in
resonance with the laser field
\begin{eqnarray}
\omega_x-\omega_y=\omega_0-m\nu_x-n\nu_y;\quad m,n\in
\mbox{\nz}\,. \label{resonance}
\end{eqnarray}
This equation can be interpreted as a condition of resonance,
which relates the difference of frequency of two Raman laser beams
to quantum numbers of two modes of vibration. The concept of the
quantum state manipulation scheme, which we present, is based on
appropriately tuned laser pulses to drive sideband transitions
with different numbers $n$,$m$ selectively. Thus, we want to
control multi-phonon transitions with simultaneous changes of the
vibrational motion in both directions. We intend to generate each
component of the quantum state (\ref{1}) from the ground state of
the collective vibration. To make this concept reliable we assume
incommensurate trap frequencies $\nu_x$ and $\nu_y$. This
requirement can be fulfilled by trap design, since the number of
photons in the quantum state (\ref{1}) is
bounded: $m\le M,n\le N$.\\
If the resonance condition (\ref{resonance}) holds for every laser
pulse the modeling can be simplified. In the example of a single
ion confined in a two-dimensional harmonic trap Steinbach
\cite{Steinbach} set a slightly different condition to describe
the phonon exchange between different directions of vibrational
motion. We do a similar kind of modelling, but with a different
goal. After the standard dipole and rotating wave approximation,
the adiabatic elimination of the auxiliary off-resonant levels
lead to an effective Hamiltonian
\begin{eqnarray}
H=H_0+H_{int} \label{3}
\end{eqnarray} with
\begin{eqnarray}
H_0=\frac{\omega_0}{2}(|g_2><g_2|-|g_1><g_1|)+\nu_xa^{\dagger}a+
\nu_yb^{\dagger}b\, ; \label{4}
\end{eqnarray}
\begin{eqnarray}
H_{int}=\Omega{e^{i\eta_x(a+a^{\dagger})+i\eta_y(b+b^{\dagger})+i(\omega_x-\omega_y)t}}|g_2><g_1|+h.c.
\,.\label{5}
\end{eqnarray}
Where $a$ and $b$ ($a^{\dagger}$ and $b^{\dagger}$) are
annihilation(creation) operators of the quantized motion along the
$x$ and $y$ axes. $\eta_x$ and $\eta_y$ are the associated
Lamb-Dicke parameters in the $x$ and $y$ direction. The complex
parameter $\Omega$ denotes the effective Raman coupling constant,
which
considers the phase difference $\phi$ of the two Raman laser beams.\\
In order to generate the motional quantum state (\ref{2}), we
consider the situation in which the ion is prepared in the ground
state $|g_2>$ and the two c.m. modes in the vacuum states $|0>_x$
and $|0>_y$:
\begin{eqnarray}
\Psi_{initial}=|g_2>|0>_x|0>_y=|g_2,0,0>\,.\label{ini}
\end{eqnarray}
In the following we show, that each term of this linear
combination can be generated from the ground
state term $|0>_x|0>_y$ most efficiently by one laser pulse.\\
In the rotating wave approximation we consider only those terms of
the operator $H_{int}$, which are in resonance with the actual
laser pulse. That is why we write the operator of interaction
conveniently in the form
\begin{eqnarray}
H_{m,n}&=&\Omega{e^{-\eta_x^2/2-\eta_y^2/2}}
\sum_{k_1,k_2=0}^{\infty}\frac{(i\eta_x)^{2k_1+m}(i\eta_y)^{2k_2+n}}{k_1!k_2!(k_1+m)!(k_2+n)!}
a^{\dagger{k_1}}a^{k_1}b^{\dagger{k_2}}b^{k_2}a^{m}b^{n}|g_2><g_1|\nonumber\\
&&+h.c. \,.\label{6}
\end{eqnarray}
The index $\{m,n\}$ indicates the relevant components of
$H_{int}$, which correspond to the condition of resonance
(\ref{resonance}). It follows, that the operator of time evolution
$U_{mn}=\exp{(-iH_{m,n}t)}$ satisfies the relations
\begin{eqnarray}
U_{mn}|g_2>|0>_x|0>_y&=&\cos(|\Omega_{m,n}|t)|g_2>|0>_x|0>_y\nonumber\\
&&-ie^{-i\phi_{m,n}}\sin(|\Omega_{m,n}|t)|g_1>|m>_x|n>_y
\end{eqnarray}
\begin{eqnarray}
U_{mn}|g_1>|k>_x|l>_y=|g_1>|k>_x|l>_y;\quad k+l<m+n
\end{eqnarray}
\begin{eqnarray}
U_{mn}|g_1>|k>_x|l>_y=|g_1>|k>_x|l>_y;\quad k+l=m+n;\, k\neq{m}\,.
\label{7}
\end{eqnarray}
Here $|\Omega_{m,n}|$ and $\phi_{m,n}$ are the amplitude and the
phase of the corresponding Raman coupling constant
\begin{eqnarray}
\Omega_{m,n}=|\Omega_{m,n}|\,e^{i\phi_{m,n}}=
\Omega{e^{-\eta_x^2/2-\eta_y^2/2+i\phi}}\frac{(i\eta_x)^{m}(i\eta_y)^{n}}{m!n!}\,
.
\end{eqnarray}
We begin from the ground state (\ref{ini}) by applying the first
laser pulse. It fulfills the resonance condition (\ref{resonance})
in the case of: $n=0$; $m=0$. This laser pulse, which corresponds
to $H_{0,0}$, has to generate the term $d_{0,0}|0,0>$ of the
quantum state (\ref{2}). We choose the duration $t_{0,0}$ of the
laser pulse to fulfill
\begin{eqnarray}
-ie^{-i\phi_{0,0}}\sin(|\Omega_{0,0}|t_{0,0})=d_{0,0} \,.\label{9}
\end{eqnarray}
The system's state is transformed into
\begin{eqnarray}
\Psi^{0,0}&=&\cos(|\Omega_{0,0}|t_{0,0})|g_2>|0,0>-ie^{-i\phi_{0,0}}\sin(|\Omega_{0,0}|t_{0,0})|g_1>|0,0>
\label{8}\\
&=&\sqrt{1-|d_{0,0}|^2}|g_2>|0,0>+d_{0,0}|g_1>|0,0> \, .\label{10}
\end{eqnarray}
In order to drive the system with the effective interaction
$H_{0,1}$ the next laser pulse is tuned according to Eq.
(\ref{resonance}) to resonance: $(m;n)=(0;1)$. After this laser
pulse has driven the ion with a duration $t_{0,1}$, the system's
state becomes
\begin{eqnarray}
\Psi^{0,1}&=&\sqrt{1-|d_{0,0}|^2}[\cos(|\Omega_{0,1}|t_{0,1})|g_2>|0,0>
\nonumber\\
&&-ie^{-i\phi_{0,1}}\sin(|\Omega_{0,1}|t_{0,1})|g_1>|\frac{1}{2},-\frac{1}{2}>]+d_{0,0}|g_1>|0,0>\,.
\label{11}
\end{eqnarray}
We choose the laser pulse duration $t_{0,1}$ to satisfy
\begin{eqnarray}
-ie^{-i\phi_{0,1}}\sqrt{1-|d_{0,0}|^2}\sin(|\Omega_{0,1}|t_{0,1})=d_{\frac{1}{2},-\frac{1}{2}}
\, .\label{12}
\end{eqnarray}
Thus, after the $(\frac{1}{2},-\frac{1}{2})$-component of Eq.
(\ref{2}) is generated, the system's state becomes
\begin{eqnarray}
\Psi^{0,1}&=&\sqrt{1-|d_{0,0}|^2-|d_{\frac{1}{2},-\frac{1}{2}}|^2}|g_2>|0,0>
\nonumber\\
&&+d_{\frac{1}{2},-\frac{1}{2}}|g_1>|\frac{1}{2},-\frac{1}{2}>+d_{0,0}|g_1>|0,0>
\, .\label{13}
\end{eqnarray}
We proceed with a laser pulse, which is characterized by
$(m;n)=(1;0)$, to let the quantum state evolve into
\begin{eqnarray}
\Psi^{1,0}&=&\sqrt{1-|d_{0,0}|^2-|d_{\frac{1}{2},-\frac{1}{2}}|^2}[\cos(|\Omega_{1,0}|t_{1,0})|g_2>|0,0>
\nonumber\\
&&-ie^{-i\phi_{1,0}}\sin(|\Omega_{1,0}|t_{1,0})|g_1>|\frac{1}{2},\frac{1}{2}>]\nonumber\\
&&+d_{\frac{1}{2},-\frac{1}{2}}|g_1>|\frac{1}{2},-\frac{1}{2}>+d_{0,0}|g_1>|0,0>\,
. \label{14}
\end{eqnarray}
With the choice
\begin{eqnarray}
-ie^{-i\phi_{1,0}}\sqrt{1-|d_{00}|^2-|d_{\frac{1}{2},-\frac{1}{2}}|^2}\sin(|\Omega_{1,0}|t_{1,0})=d_{\frac{1}{2},\frac{1}{2}}
\label{15}
\end{eqnarray} we obtain
\begin{eqnarray}
\Psi^{1,0}&=&\sqrt{1-|d_{0,0}|^2-|d_{\frac{1}{2},-\frac{1}{2}}|^2-|d_{\frac{1}{2},\frac{1}{2}}|^2}|g_2>|0,0>
\nonumber\\
&&+d_{\frac{1}{2},\frac{1}{2}}|g_1>|\frac{1}{2},\frac{1}{2}>+d_{\frac{1}{2},-\frac{1}{2}}|g_1>|\frac{1}{2},-\frac{1}{2}>+d_{00}|g_1>|0,0>
\,.\label{16}
\end{eqnarray}
If this procedure is done for the
$[\frac{(i+j)(i+j+1)}{2}+i+1]-$th time, the quantum state of the
system is
\begin{eqnarray}
\Psi^{i,j}&=&\sum_{J=0}^{(i+j-1)/2}\sum_{L=-J}^{J}d_{J,L}|J,L>|g_1>+\sum_{L=-(i+j)/2}^{(i-j)/2}d_{(i+j)/2,L}|(i+j)/2,L>|g_1>
\nonumber\\
&&+\sqrt{1-\sum_{J=0}^{(i+j-1)/2}\sum_{L=-J}^{J}|d_{J,L}|^2-\sum_{L=-(i+j)/2}^{(i-j)/2}|d_{(i+j)/2,L}|^2}|g_2>|0,0>
\,.\label{17}
\end{eqnarray}
We now consider the $[\frac{(i+j)(i+j+1)}{2}+i+2]-$th operation by
discussing the two possible cases.\\
In the special case of $j=0$ we drive the ion with the operator of
interaction $H_{0,i+1}$. After the system has been driven by the
corresponding laser pulse of duration $t_{0,i+1}$, the quantum
state becomes
\begin{eqnarray}
\Psi^{0,i+1}&=&\sum_{J=0}^{(i)/2}\sum_{L=-J}^{J}d_{J,L}|J,L>|g_1>\nonumber\\
&&+\sqrt{1-\sum_{J=0}^{(i)/2}\sum_{L=-J}^{J}|d_{J,L}|^2}[\cos(|\Omega_{0,i+1}|t_{0,i+1})|g_2>|0,0>
\nonumber\\
&&-ie^{-i\phi_{0,i+1}}\sin(|\Omega_{0,i+1}|t_{0,i+1})|g_1>|(i+1)/2,-(i+1)/2>]
\, .\label{18}
\end{eqnarray}
With the choice
\begin{eqnarray}
-ie^{-i\phi_{0,i+1}}\sqrt{1-\sum_{J=0}^{(i)/2}\sum_{L=-J}^{J}|d_{J,L}|^2}\sin(|\Omega_{0,i+1}|t_{0,i+1})=d_{(i+1)/2,-(i+1)/2}
\label{19}
\end{eqnarray} we obtain
\begin{eqnarray}
\Psi^{0,i+1}&=&\sum_{J=0}^{(i)/2}\sum_{L=-J}^{J}d_{J,L}|J,L>|g>+d_{(i+1)/2,-(i+1)/2}|(i+1)/2,-(i+1)/2>|g_1>
\nonumber\\
&&+\sqrt{1-\sum_{J=0}^{(i)/2}\sum_{L=-J}^{J}|d_{J,L}|^2-|d_{(i+1)/2,-(i+1)/2}|^2}|g_2>|0,0>
\,.\label{20}
\end{eqnarray}
In the other case ($j\neq{0}$) the ion is driven by a laser pulse,
which corresponds to $H_{i+1,j-1}$. We choose the interaction time
$t_{i+1,j-1}$ and phase of laser field to satisfy
\begin{eqnarray}
d_{(i+j)/2,(i-j+2)/2}&=&-ie^{-i\phi_{i+1,j-1}}\sin(|\Omega_{i+1,j-1}|t_{i+1,j-1})
\nonumber\\
&&\times\sqrt{1-\sum_{J=0}^{(i+j-1)/2}\sum_{L=-J}^{J}|d_{J,L}|^2-\sum_{L=-(i+j)/2}^{(i-j)/2}|d_{(i+j)/2L}|^2}
 \label{21}
\end{eqnarray}
and to obtain the quantum state
\begin{eqnarray}
\Psi^{i+1,j-1}&=&\sum_{J=0}^{(i+j-1)/2}\sum_{L=-J}^{J}d_{J,L}|J,L>|g_1>+\sum_{L=-(i+j)/2}^{(i-j+2)/2}d_{(i+j)/2,L}|(i+j)/2,L>|g_1>
\nonumber\\
&&+\sqrt{1-\sum_{J=0}^{(i+j-1)/2}\sum_{L=-J}^{J}|d_{J,L}|^2-\sum_{L=-(i+j)/2}^{(i-j+2)/2}|d_{(i+j)/2,L}|^2}|g_2>|0,0>
\, .\label{22}
\end{eqnarray}
After the procedure is performed for $\frac{(M+N+1)(M+N)}{2}$
times the system's state definitely becomes
\begin{eqnarray}
\Psi_{final}=\sum_{J=0}^{(M+N)/2}\sum_{L=-J}^{J}d_{J,L}|J,L>|g_1>
\, . \label{23}
\end{eqnarray}
Thus, the system is prepared in a product state $\Psi$ of the
desired vibrational quantum state (2) and the ground state
$|g_1>$.\\

We have proposed a scheme to generate any two-dimensional quantum
state of vibration of one trapped ion. This concept is based on
the possibility to drive sideband transitions with different
numbers $n$, $m$ selectively by tuning laser pulses appropriately.
Thus, we want to control multi-phonon transitions with
simultaneous changes of the quantum numbers of vibration in both
directions. Each component of the quantum state (\ref{1}) can be
generated step by step from the ground state of vibration. To make
our concept reliable we have assumed that the trap frequencies
$\nu_x$ and $\nu_y$ can be made incommensurate by trap design.

However, there might exist on-resonant terms in addition to the
ones included in Eq. (8) due to finite bandwith of laser pulse. If
the ratio of the trap frequencies is chosen appropriately so that
frequency of each lase pulse is enough separated,  these terms can
be neglected in the Lamb-Dicke limit. We assume equal values of
the Lamb-Dicke parameter $\eta_x=\eta_y$\cite{Steinbach} and
require a ratio of trap frequencies to satisfy condition
$\nu_x/\nu_y>M+2N$. In this csae, we see that the laser frequency
of each sideband transition is enough separated and we can neglect
those unwanted resonant terms. Thus, in order to implement our
scheme, we require a large enough trap anisotropies. In addition,
in Lamb-Dicke limit, the effective Rabi frequency $\Omega_{m,n}$
of Eq. (12) is proportional to $
\frac{(i\eta_x)^{m}(i\eta_y)^{n}}{m!n!}$. If the quantum numbers
of vibration $n$ and $m$ are too large the effective Rabi
frequency $\Omega_{m,n}$ becomes too small. Thus, in respect of
decoherence the time required to complete the procedure might
become too long. This problem represents the
limitation of this scheme.\\
This scheme of quantum state generation requires not more than
(M+1)(N+1) laser pulses, which is the smallest possible number.
Thus, in respect of short laser pulse sequences it can be
considered as the optimal solution of this problem of the quantum
state generation. The number of required laser pulses reduces, if
there are coefficients in the target quantum state, which are
equal to zero. For example, if the coefficient $d_{J,L}$ of
Eq.(\ref{2}), which is generated with the $(2J^2+J+L+1)$-th
operation, is zero, it is not necessary to apply the corresponding
laser pulse.

\end{document}